\RequirePackage{lineno}
\documentclass[aps,prl,twocolumn,showpacs,superscriptaddress,groupedaddress,floatfix]{revtex4}  
\usepackage{graphicx}  
\usepackage{dcolumn}   
\usepackage{bm}        
\usepackage{amssymb}   
\usepackage{color} 
\usepackage{xspace} 

\newcommand{\sigeff}{$\sigma_{\rm eff}$}

\newcommand{\ptjs}{$p^{J/\psi}_T$}

\begin{document}
\widetext

 \hspace{5.2in} \mbox{FERMILAB-PUB-15-485-E}
\title{Evidence for Simultaneous Production of $J/\psi$ and $\Upsilon$ Mesons}
 \affiliation{LAFEX, Centro Brasileiro de Pesquisas F\'{i}sicas, Rio de Janeiro, Brazil}
\affiliation{Universidade do Estado do Rio de Janeiro, Rio de Janeiro, Brazil}
\affiliation{Universidade Federal do ABC, Santo Andr\'e, Brazil}
\affiliation{University of Science and Technology of China, Hefei, People's Republic of China}
\affiliation{Universidad de los Andes, Bogot\'a, Colombia}
\affiliation{Charles University, Faculty of Mathematics and Physics, Center for Particle Physics, Prague, Czech Republic}
\affiliation{Czech Technical University in Prague, Prague, Czech Republic}
\affiliation{Institute of Physics, Academy of Sciences of the Czech Republic, Prague, Czech Republic}
\affiliation{Universidad San Francisco de Quito, Quito, Ecuador}
\affiliation{LPC, Universit\'e Blaise Pascal, CNRS/IN2P3, Clermont, France}
\affiliation{LPSC, Universit\'e Joseph Fourier Grenoble 1, CNRS/IN2P3, Institut National Polytechnique de Grenoble, Grenoble, France}
\affiliation{CPPM, Aix-Marseille Universit\'e, CNRS/IN2P3, Marseille, France}
\affiliation{LAL, Universit\'e Paris-Sud, CNRS/IN2P3, Orsay, France}
\affiliation{LPNHE, Universit\'es Paris VI and VII, CNRS/IN2P3, Paris, France}
\affiliation{CEA, Irfu, SPP, Saclay, France}
\affiliation{IPHC, Universit\'e de Strasbourg, CNRS/IN2P3, Strasbourg, France}
\affiliation{IPNL, Universit\'e Lyon 1, CNRS/IN2P3, Villeurbanne, France and Universit\'e de Lyon, Lyon, France}
\affiliation{III. Physikalisches Institut A, RWTH Aachen University, Aachen, Germany}
\affiliation{Physikalisches Institut, Universit\"at Freiburg, Freiburg, Germany}
\affiliation{II. Physikalisches Institut, Georg-August-Universit\"at G\"ottingen, G\"ottingen, Germany}
\affiliation{Institut f\"ur Physik, Universit\"at Mainz, Mainz, Germany}
\affiliation{Ludwig-Maximilians-Universit\"at M\"unchen, M\"unchen, Germany}
\affiliation{Panjab University, Chandigarh, India}
\affiliation{Delhi University, Delhi, India}
\affiliation{Tata Institute of Fundamental Research, Mumbai, India}
\affiliation{University College Dublin, Dublin, Ireland}
\affiliation{Korea Detector Laboratory, Korea University, Seoul, Korea}
\affiliation{CINVESTAV, Mexico City, Mexico}
\affiliation{Nikhef, Science Park, Amsterdam, the Netherlands}
\affiliation{Radboud University Nijmegen, Nijmegen, the Netherlands}
\affiliation{Joint Institute for Nuclear Research, Dubna, Russia}
\affiliation{Institute for Theoretical and Experimental Physics, Moscow, Russia}
\affiliation{Moscow State University, Moscow, Russia}
\affiliation{Institute for High Energy Physics, Protvino, Russia}
\affiliation{Petersburg Nuclear Physics Institute, St. Petersburg, Russia}
\affiliation{Instituci\'{o} Catalana de Recerca i Estudis Avan\c{c}ats (ICREA) and Institut de F\'{i}sica d'Altes Energies (IFAE), Barcelona, Spain}
\affiliation{Uppsala University, Uppsala, Sweden}
\affiliation{Taras Shevchenko National University of Kyiv, Kiev, Ukraine}
\affiliation{Lancaster University, Lancaster LA1 4YB, United Kingdom}
\affiliation{Imperial College London, London SW7 2AZ, United Kingdom}
\affiliation{The University of Manchester, Manchester M13 9PL, United Kingdom}
\affiliation{University of Arizona, Tucson, Arizona 85721, USA}
\affiliation{University of California Riverside, Riverside, California 92521, USA}
\affiliation{Florida State University, Tallahassee, Florida 32306, USA}
\affiliation{Fermi National Accelerator Laboratory, Batavia, Illinois 60510, USA}
\affiliation{University of Illinois at Chicago, Chicago, Illinois 60607, USA}
\affiliation{Northern Illinois University, DeKalb, Illinois 60115, USA}
\affiliation{Northwestern University, Evanston, Illinois 60208, USA}
\affiliation{Indiana University, Bloomington, Indiana 47405, USA}
\affiliation{Purdue University Calumet, Hammond, Indiana 46323, USA}
\affiliation{University of Notre Dame, Notre Dame, Indiana 46556, USA}
\affiliation{Iowa State University, Ames, Iowa 50011, USA}
\affiliation{University of Kansas, Lawrence, Kansas 66045, USA}
\affiliation{Louisiana Tech University, Ruston, Louisiana 71272, USA}
\affiliation{Northeastern University, Boston, Massachusetts 02115, USA}
\affiliation{University of Michigan, Ann Arbor, Michigan 48109, USA}
\affiliation{Michigan State University, East Lansing, Michigan 48824, USA}
\affiliation{University of Mississippi, University, Mississippi 38677, USA}
\affiliation{University of Nebraska, Lincoln, Nebraska 68588, USA}
\affiliation{Rutgers University, Piscataway, New Jersey 08855, USA}
\affiliation{Princeton University, Princeton, New Jersey 08544, USA}
\affiliation{State University of New York, Buffalo, New York 14260, USA}
\affiliation{University of Rochester, Rochester, New York 14627, USA}
\affiliation{State University of New York, Stony Brook, New York 11794, USA}
\affiliation{Brookhaven National Laboratory, Upton, New York 11973, USA}
\affiliation{Langston University, Langston, Oklahoma 73050, USA}
\affiliation{University of Oklahoma, Norman, Oklahoma 73019, USA}
\affiliation{Oklahoma State University, Stillwater, Oklahoma 74078, USA}
\affiliation{Oregon State University, Corvallis, Oregon 97331, USA}
\affiliation{Brown University, Providence, Rhode Island 02912, USA}
\affiliation{University of Texas, Arlington, Texas 76019, USA}
\affiliation{Southern Methodist University, Dallas, Texas 75275, USA}
\affiliation{Rice University, Houston, Texas 77005, USA}
\affiliation{University of Virginia, Charlottesville, Virginia 22904, USA}
\affiliation{University of Washington, Seattle, Washington 98195, USA}
\author{V.M.~Abazov} \affiliation{Joint Institute for Nuclear Research, Dubna, Russia}
\author{B.~Abbott} \affiliation{University of Oklahoma, Norman, Oklahoma 73019, USA}
\author{B.S.~Acharya} \affiliation{Tata Institute of Fundamental Research, Mumbai, India}
\author{M.~Adams} \affiliation{University of Illinois at Chicago, Chicago, Illinois 60607, USA}
\author{T.~Adams} \affiliation{Florida State University, Tallahassee, Florida 32306, USA}
\author{J.P.~Agnew} \affiliation{The University of Manchester, Manchester M13 9PL, United Kingdom}
\author{G.D.~Alexeev} \affiliation{Joint Institute for Nuclear Research, Dubna, Russia}
\author{G.~Alkhazov} \affiliation{Petersburg Nuclear Physics Institute, St. Petersburg, Russia}
\author{A.~Alton$^{a}$} \affiliation{University of Michigan, Ann Arbor, Michigan 48109, USA}
\author{A.~Askew} \affiliation{Florida State University, Tallahassee, Florida 32306, USA}
% ...
%
%
\collaboration{The D0 Collaboration\footnote{with visitors from
$^{a}$Augustana College, Sioux Falls, SD, USA,
$^{b}$The University of Liverpool, Liverpool, UK,
$^{c}$DESY, Hamburg, Germany,
$^{d}$CONACyT, Mexico City, Mexico,
$^{e}$SLAC, Menlo Park, CA, USA,
$^{f}$University College London, London, UK,
$^{g}$Centro de Investigacion en Computacion - IPN, Mexico City, Mexico,
$^{h}$Universidade Estadual Paulista, S\~ao Paulo, Brazil,
$^{i}$Karlsruher Institut f\"ur Technologie (KIT) - Steinbuch Centre for Computing (SCC),
D-76128 Karlsruhe, Germany,
$^{j}$Office of Science, U.S. Department of Energy, Washington, D.C. 20585, USA,
$^{k}$American Association for the Advancement of Science, Washington, D.C. 20005, USA,
$^{l}$Kiev Institute for Nuclear Research, Kiev, Ukraine,
$^{m}$University of Maryland, College Park, MD 20742, USA,
$^{n}$European Orgnaization for Nuclear Research (CERN), Geneva, Switzerland
and
$^{o}$Purdue University, West Lafayette, IN 47907, USA.
}} \noaffiliation
\vskip 0.25cm
 \date{February 1, 2016}

\begin{abstract}
We report evidence for the simultaneous production of $J/\psi$  and $\Upsilon$ mesons in 8.1 fb$^{-1}$ of data 
collected  at $\sqrt{s}=$1.96~TeV  by the D0 experiment at the Fermilab $p \bar p$ Tevatron Collider.
Events with these characteristics are expected to be produced predominantly by gluon-gluon interactions.
In this analysis, we extract the effective cross section characterizing the initial parton spatial distribution,   
\sigeff~$=2.2\pm 0.7\mbox{\thinspace(stat)} \pm 0.9 \mbox{\thinspace(syst)} ~\mbox{mb}$. 
\end{abstract}

\maketitle

The importance of multiple parton interactions (MPI) in hadron-hadron collisions 
as a background to processes such as Higgs production or various new phenomena has been often underestimated in the past.
For instance, in the associated production of Higgs and weak bosons, where the Higgs boson decays into $b \bar b$, the MPI background, 
in which one interaction produces the vector boson and another produces a pair of jets,
 may exceed the size of the Higgs signal
even after the application of strict event selections~\cite{Bandurin:2010gn}.
Recent data~\cite{cdf2,atlas,D0_2,cms_w2j,atlasjj,D0JJ,LHCb,CMSJJ} examining various double parton interactions have attracted considerable theoretical attention~\cite{JJ_LHC,barsnigzot,Bandurin:2010gn,jpsiprd,newjy,newjjls}.

In this Letter, we measure for the first time the cross section for simultaneous production of $J/\psi$ and $\Upsilon$ ($1S,2S,3S$) mesons in $p \bar p$ collisions at $\sqrt{s} = 1.96$ TeV.
   The production of two quarkonium states can be used to probe the interplay of perturbative and nonperturbative phenomena in quantum
 chromodynamics (QCD) and to search for new bound states of hadronic matter such as tetraquarks~\cite{JJ_LHC,humpert}. 
 Here we focus on double quarkonium production as a measure of the spatial distribution of partons in the nucleon.

Unlike other quarkonium processes such as double $J/\psi$ production, or processes involving jets or vector bosons, the production  of $J/\psi$ and $\Upsilon$ mesons is expected to be dominated by double parton (DP) interactions involving the collisions
 of two independent pairs of partons within the colliding beam particles.
 The simultaneous production through single parton (SP) interactions
 is suppressed by additional powers of $\alpha_s$ and by the small size of the allowed color octet matrix elements~\cite{barsnigzot}.
   The DP process is estimated in Ref.~\cite{newjy} to give the dominant contribution to the total $J/\psi + \Upsilon$ production at the Tevatron. In this analysis, we assume that there is no SP contribution~\cite{new_ref_dp}.
Because of the dominance of $gg$ interactions in producing heavy quarkonium states, the spatial distribution of gluons in a proton~\cite{Dok,vant,barsnigzotnew}
is directly probed by the DP scattering rate, which represents simultaneous, independent parton interactions.
In contrast, the DP studies involving vector bosons and jets probe the spatial distributions of quark-quark or quark-gluon initial states~\cite{cdf2,D0_2,atlas,cms_w2j,atlasjj}.
 
In $p\bar{p}$ collisions, there are three main production mechanisms for $J/\psi$ mesons: 
prompt production;
as a radiative decay product of
promptly produced heavier charmonium states such as the 
$^3{}\!P_1$ state $\chi_{1c}$ and the $^3{}\!P_2$ state $\chi_{2c}$; and 
 nonprompt $B$ hadron decays. 
A particle is considered produced promptly if it originates in the initial $p \bar p$ interaction or if it originates in either an electromagnetic or strong force mediated decay
 and thus the tracks appear to be produced at the $p \bar p$ interaction vertex.
$\Upsilon$ mesons are only produced promptly, either directly or as decay products of higher mass states, such as $\chi_{1b}$ or $\chi_{2b}$. 
Prompt heavy quarkonium production is described by three types of models:
the color-singlet (CS) model~\cite{color_singlet}; the color evaporation model~\cite{color_evaporation,color_evaporation_1} with a subsequent soft color interaction model~\cite{color_interaction};
and the color-octet (CO) model~\cite{color_octet,color_octet_1}.

In this Letter, we present the first measurement of the cross section of the simultaneous production of prompt $J/\psi$ and $\Upsilon$ mesons, 
 as well as
a measurement of the single prompt $J/\psi$ production cross section.
The $\Upsilon$ cross section was measured previously by D0~\cite{d0_ups}.
The measurements are based on a data sample collected
by the D0 experiment at the Tevatron 
corresponding to an integrated luminosity of $8.1\pm0.5$~fb$^{-1}$~\cite{d0lum}. 
Assuming that the simultaneous production of $J/\psi$ and $\Upsilon$ mesons is caused solely by DP scattering,
we extract the effective cross section~($\sigma_{\rm{eff}}$), 
a parameter related to an initial state parton spatial density
distribution within a nucleon (see, e.g., Ref.~\cite{barsnigzotnew}):
\begin{eqnarray}
\sigma_{\rm eff}^{-1} = \int d^2\beta [F({\bf \beta})]^2
\end{eqnarray}
with $F({\bf \beta})=\int f({\bf b}) f({\bf b}-{\bf \beta})d^2b$,
where ${\bf \beta}$ is the vector impact parameter of the two colliding hadrons,
and $f({\bf b})$ is a function describing the transverse spatial distribution
of the partonic matter inside a hadron.
The $f({\bf b})$ may depend on the parton flavor.

The cross section for double parton scattering, $\sigma_{\rm DP}$, is related to $\sigma_{\rm eff}$ 
for the  production  of $J/\psi$ and $\Upsilon$ mesons:
\begin{equation}
\sigma_{\rm{eff}} = \frac{\sigma(J/\psi)\sigma(\Upsilon)}{\sigma_{\mathrm {DP}}(J/\psi+\Upsilon)}.
\label{eq:s_eff}
\end{equation}
Both the $J/\psi$ and $\Upsilon$ mesons are fully reconstructed via their decay $J/\psi(\Upsilon) \rightarrow \mu^+ \mu^-$, where
 the muons are required to have  transverse momenta $p^\mu_T>2$ GeV/c and pseudorapidity $|\eta^\mu|<2.0$~\cite{coord}.
The cross sections measured with these kinematic requirements are referred to below as "fiducial cross sections."

The general purpose D0 detector 
is described in detail elsewhere~\cite{D0det,D0det1}. 
The two subdetectors used to trigger and reconstruct muon final states are the muon and the central tracking systems.
The central tracking system, used to reconstruct charged particle tracks, consists of the inner silicon microstrip tracker (SMT)~\cite{smt}
and outer central fiber tracker (CFT) detector both placed inside a 1.9~T solenoidal magnet. The solenoidal magnet is located  inside  the central calorimeter. 
The muon detectors~\cite{RunI_muon} surrounding the calorimeters
consist of three layers of drift tubes and three layers of scintillation counters, one inside the 1.8~T iron toroidal magnets and
two outside. The luminosity is measured using plastic scintillation counters surrounding the beams at small polar angles~\cite{d0lum}.

We require  events to pass at least 
one of a set of low-$p_T$ dimuon triggers. 
The identification of muons starts with requiring hits at least in the muon detector layer 
in front of the toroids~\cite{RunII_muon} and proceeds by
matching the hits in the muon system to a charged particle track reconstructed by the central tracking system. The track is required to have at least one hit in the SMT and at least two hits in the CFT detectors.
To suppress cosmic rays, the muon candidates must satisfy strict timing requirements.
 Their distance of the closest approach to the beam line
has to be less than $0.5$ cm and their matching tracks have to pass within 2 cm 
of the primary $p \bar p$ interaction vertex along the beam axis. 
We require two oppositely charged muons, isolated in the calorimeter and tracking detectors~\cite{RunII_muon},
 with good matching of the tracks in the inner tracking and those in the muon detector,
 and masses within the ranges $2.4 < M_{\mu \mu} < 4.2$ GeV or $8 < M_{\mu \mu} < 12$ GeV for the $J/\psi$ and $\Upsilon$ candidates, respectively.
 The mass windows are chosen to be large enough to provide an estimate of backgrounds on either side of the $J/\psi$ or $\Upsilon$ mass peaks. 
Events that have a pair of such muons in each of the two invariant mass windows are identified as $J/\psi$ and $\Upsilon$ simultaneous production candidates. 
Background events are mainly due to random combinations of muons from $\pi^{\pm}$, $K^{\pm}$ decays (decay background),
 continuous nonresonant $\mu^{+}\mu^{-}$ Drell-Yan (DY) production,  and  $B$ hadron  decays into $J/\psi+X$.
 In the case of $J/\psi+\Upsilon$ production, there is also a background where one muon pair results from a genuine $J/\psi$ or $\Upsilon$ decay and the other pair is a nonresonant combination of muons ($J/\psi(\Upsilon)+\mu \mu$). 

In our single quarkonium sample, the backgrounds from $\pi^{\pm}$, $K^{\pm}$ decays and DY events are estimated simultaneously with the number of signal events by 
performing a fit to the $M_{\mu \mu}$ invariant mass distribution using a superposition of Gaussian functions for signal and a quadratic function for the background. 
The $\psi(2S)$ events are included in the fitted region but omitted for the single $J/\psi$ cross section calculation, while all three $\Upsilon$ mass states ($1S,2S,3S$) are included in the $\Upsilon$ cross section calculation.
The number of single $J/\psi$ events found in the fit is $6.9\times 10^6$, while the number of
single $\Upsilon$ events is $2.1\times 10^6$. 

The single $J/\psi$  trigger efficiency is estimated using events with a reconstructed $J/\psi$ which pass
zero-bias (ZB) triggers requiring only a beam crossing, or minimum bias (MB) triggers
which  only require  hits in the luminosity detectors, and  that do or do not satisfy the dimuon trigger requirement.
To estimate the trigger efficiency for the $\Upsilon$ selection, we use the $\Upsilon$(1S) cross section previously measured by the D0 experiment~\cite{d0_ups}, extrapolated to our fiducial region using
events generated with the {\sc pythia}~\cite{pythia} Monte Carlo (MC) event generator
and increased to include the $\Upsilon$(2S, 3S) contributions.
 Using {\sc pythia} for the extrapolation introduces a negligible bias because
the fiducial regions are similar and the D0 muon system acceptance outside both fiducial regions is low.
The trigger efficiencies for single $J/\psi$ mesons and for single $\Upsilon$ mesons in the fiducial region are $0.13 \pm 0.03$\thinspace(syst)  and $0.29 \pm 0.05$\thinspace(syst), respectively, where the systematic uncertainties are dominated by the small size of the ZB and MB samples. 
 The trigger efficiency for the $J/\psi+\Upsilon$ selection is estimated using the single $J/\psi$ and $\Upsilon$ trigger efficiencies and MC
samples of $J/\psi+\Upsilon$ events generated with the {\sc pythia} MC generator. 
The events are passed  through a {\sc geant} based~\cite{geant} simulation of the D0 detector and  overlaid  with data  ZB events to mimic event pileup, and processed with the same reconstruction software  as data.
We calculate the trigger efficiency for every possible pairing of muons in DP $J/\psi+\Upsilon$ MC events using the parametrizations of the dimuon trigger efficiencies as a function of $p_T^{J/\psi}$ and $p_T^{\Upsilon}$ and obtain an efficiency of  $0.77 \pm 0.04$\thinspace(syst). The substantial increase in the trigger efficiency is due to the presence of four muons in the $J/\psi+\Upsilon$ events.

We use {\sc pythia}-generated single $J/\psi$ and $\Upsilon$ events to estimate  the combined geometric and kinematic acceptance and
reconstruction efficiency.
The  generated and reconstructed events  are selected using the same muon selection criteria.
We correct the number of simulated reconstructed events for the different reconstruction efficiencies in data and MC events, calculated
in ($p_T^\mu$, $\eta^\mu$) bins.
The product of the acceptance and efficiency for 
single $J/\psi$ events produced in the color singlet model is 
$0.19\pm0.01\thinspace({\rm syst})$.
The product of the acceptance and efficiency for 
single $\Upsilon$ events  is 
$0.43\pm0.05\thinspace({\rm syst})$.
The systematic uncertainties are due to muon identification efficiency mismodeling and to the differences in the kinematic distributions 
between the data and simulated $J/\psi$ or $\Upsilon$ events.
 The $\cos \theta^*$ distribution, 
 where $\theta^*$ is the polar angle of the decay muon in the Collins-Soper frame~\cite{CS}, is sensitive to the $J/\psi$ and $\Upsilon$ polarizations~\cite{baranov_polar,CDF_pol,CMS_pol,cdf_ups_pol,cms_ups_pol}.   
Data-to-MC reweighting factors based on the observed $\cos \theta^*$ distribution are used to recalculate the acceptance,
  and lead to $\lesssim1\%$ difference  with the default acceptance value for single $J/\psi$ events and $\approx11\%$ for single $\Upsilon$ events, which we take as systematic uncertainties. 

The vertex of a $B$ hadron decay
into the $J/\psi+X$ final state is on average several hundred microns away from 
the $p\bar{p}$ interaction vertex, while prompt $J/\psi$ production occurs directly at
 the  interaction point. 
To identify promptly produced $J/\psi$ mesons, 
we examine  the decay length from the primary $p\bar{p}$ interaction vertex (in the plane transverse to the beam)  to the $J/\psi$ production vertex,
defined as $c \tau = L_{xy} m_{J/\psi}/p^{J/\psi}_T$, where $L_{xy}$ is
calculated as the distance between the intersection of the muon tracks and the $p \bar p$ interaction vertex, $m_{J/\psi}$
is the world average $J/\psi$ mass~\cite{PDG},
and $p^{J/\psi}_T$ is the $J/\psi$ transverse momentum.

\begin{figure}[h!]
\includegraphics[width=0.47\textwidth,keepaspectratio=true]{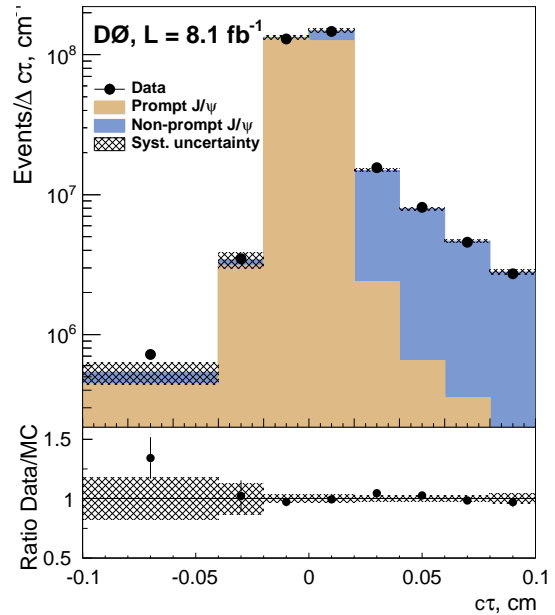}
\caption{
The $c \tau$ distribution of background subtracted single $J/\psi$ events
after all selection criteria. The distributions for the
signal and background templates are shown normalized to
their respective fitted fractions with $\chi^2/\rm{Ndof}=1.6$, $\rm{Ndof}=6$. 
The shaded uncertainty band corresponds to the total systematic uncertainty on the sum of signal and background events. 
}
\label{fig:single_jp}
\end{figure}

The fraction of prompt $J/\psi$ mesons in the data sample is estimated by performing a maximum likelihood fit 
of the $c \tau$ distribution. The fit uses templates 
for the prompt $J/\psi$ signal events,
taken from the single $J/\psi$ MC sample, and for nonprompt $J/\psi$ events,
taken from  the $b \bar b$ MC sample.
Both are generated with {\sc pythia}.
The prompt $J/\psi$ fraction obtained from the fit is $0.83~\pm~0.03$\thinspace(syst). 
The systematic uncertainty is dominated by the uncertainty in the MC modeling of the $c \tau$.
The fit result is shown in Fig.~\ref{fig:single_jp}.
By applying the selection $c \tau<0.02~(>0.03)$ cm, we verify that the \ptjs spectra of the  prompt (nonprompt) $J/\psi$ events in data are well described by MC
simulations in the prompt ($B$-decay) dominated regions.

The fiducial cross section of the prompt single $J/\psi$ production is calculated 
using the number of $J/\psi$ candidates in data, the fraction of prompt $J/\psi$ events, the trigger efficiency,
the acceptance and selection efficiencies, as well as the integrated luminosity.
The fiducial cross section is
\begin{eqnarray}
 \sigma(J/\psi) = 28 \pm 7 \mbox{\thinspace(syst)} ~\mbox{nb}.
 \label{eq:CS2_sj}
\end{eqnarray}
The systematic uncertainty in the single $J/\psi$ cross section mainly arises from the trigger efficiency. The statistical uncertainty is negligible.
The measured single $J/\psi$ cross section is in agreement with the measurement by D0~\cite{D0JJ} [$23.9\pm4.6\thinspace({\rm stat})\pm3.7\thinspace({\rm syst})$ nb] in a similar fiducial region and with
the measurement by CDF~\cite{cdf_single} if an interpolation to the CDF fiducial region is performed.

The cross section for single $\Upsilon$ production is extrapolated to our fiducial region from the previous D0 measurement~\cite{d0_ups}.
Using the ratio of $\Upsilon$(1S)
to $\Upsilon$(sum of $1S,2S,3S$ states)
of $0.73 \pm 0.03\mbox{\thinspace(syst)}$, estimated in $\Upsilon$ selection data, we obtain the $\Upsilon$ cross section (the statistical uncertainty is  negligible):
\begin{eqnarray}
\sigma(\Upsilon) = 2.1 \pm 0.3 \mbox{\thinspace(syst)} ~\mbox{nb}.
 \label{eq:CS2_sy}
\end{eqnarray}
The systematic uncertainty in $\sigma(\Upsilon)$ includes that from Ref.~\cite{d0_ups} as well as those from the $\Upsilon$(1S) fraction and the extrapolation to the fiducial region.

\begin{figure}[htb]
\includegraphics[width=0.5\textwidth,keepaspectratio=true]{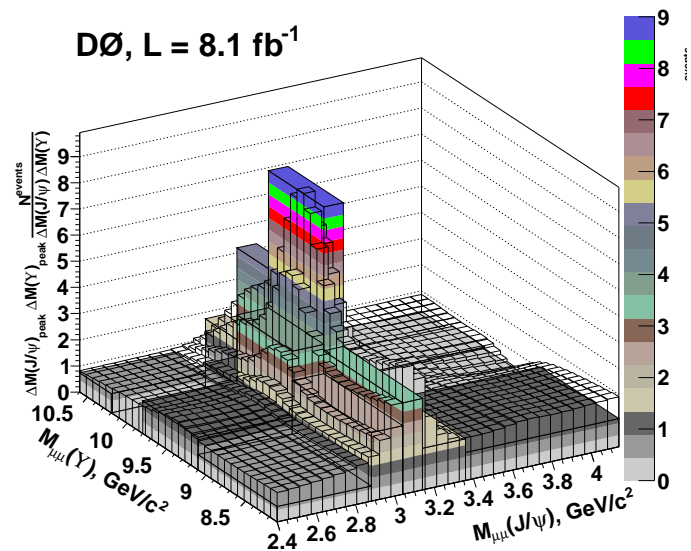} 
  \caption{ Dimuon invariant mass distribution in data  for two muon pairs, $M_{\mu\mu}(J/\psi)$, $M_{\mu\mu}(\Upsilon)$, divided by the bin area, after the  selection criteria. 
    Also shown is the two-dimensional fit surface. 
The factor $\Delta M(J/\psi)_{\rm peak}$ $\Delta M(\Upsilon)_{\rm peak}$ is applied so that the height of the peak bin is the number of observed events in that bin.}
  \label{fig:mass_2D}
\end{figure}

\begin{figure}[htb]
\includegraphics[width=0.5\textwidth,keepaspectratio=true]{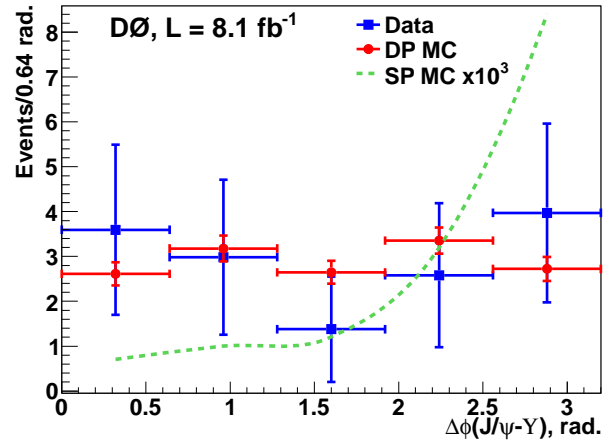} 
  \caption{The distribution of the azimuthal angle between the $J/\psi$ and $\Upsilon$ candidates, $\Delta \phi(J/\psi, \Upsilon)$, in data
 after background subtraction, in DP MC~\cite{pythia}, and SP MC~\cite{baranov} results. MC events are in arbitrary units.}
  \label{fig:dydphi}
\end{figure}

In the data, 21 events pass the selection criteria for $J/\psi+\Upsilon$ pair production in the $J/\psi$ mass window $2.88 < M_{\mu \mu} < 3.36$ GeV$/c^2$ and $\Upsilon$ mass window $9.1 < M_{\mu \mu} < 10.2$ GeV$/c^2$.
Figure~\ref{fig:mass_2D} shows the distribution of the two dimuon masses [$M_{\mu\mu}(J/\psi,\Upsilon)$] in these
and surrounding mass regions. We estimate the accidental and $J/\psi(\Upsilon)+\mu\mu$ backgrounds using the same technique
 of combining the one-dimensional functional forms utilized in single $J/\psi$ and $\Upsilon$ signal and background parametrizations as in Ref.~\cite{D0JJ}.
We fit a two-dimensional distribution of the $M_{\mu\mu}(J/\psi,\Upsilon)$ with the resulting two-dimensional functional form
 and estimate the number of $J/\psi+\Upsilon$ events is $14.5 \pm 4.6\thinspace({\rm stat}) \pm 3.4\thinspace({\rm syst})$.
This corresponds to a prompt $J/\psi+\Upsilon$ signal of $12.0 \pm 3.8\thinspace({\rm stat}) \pm 2.8\thinspace({\rm syst})$ events. 
The probability of the observed number of events to have arisen from the background is $6.3 \times 10^{-4}$, corresponding to 3.2 standard deviation evidence for the production of prompt $J/\psi+\Upsilon$.
The probability calculation includes the systematic uncertainties in the background estimates.
The distribution of the azimuthal angle between the $J/\psi$ and $\Upsilon$ candidates, $\Delta \phi(J/\psi, \Upsilon)$ 
after the subtraction of backgrounds
is shown in Fig.~\ref{fig:dydphi}. The data distribution is consistent with the DP MC model, which is uniform~\cite{barsnigzot}, substantiating our assumption 
that the DP process is the dominant contribution to the selected $J/\psi+\Upsilon$ data sample. 

We estimate the acceptance, reconstruction, and selection efficiencies for  $J/\psi+\Upsilon$ events using MC DP samples. 
 The product of the acceptance and the selection  efficiency for the DP events is found to be  
$(A \varepsilon_{s})~=~0.071~\pm~0.007\thinspace{\rm (syst)}$,
where the systematic uncertainty is dominated by the uncertainty in the modeling of the $J/\psi$ and $\Upsilon$ kinematics and muon identification efficiency for our sample with low $p_T$ muons.

Using the numbers presented above, we obtain the cross section of the simultaneous production of $J/\psi$ and $\Upsilon$ mesons:
\begin{eqnarray}
\sigma_{\rm{DP}}(J/\psi+\Upsilon)=27 \pm 9 \mbox{\thinspace(stat)} \pm 7 \mbox{\thinspace(syst)} ~\mbox{fb}.
 \label{eq:CS2_djj_DP}
\end{eqnarray}

From the measured cross sections of prompt single $J/\psi$, DP $J/\psi+\Upsilon$, and the estimate of single $\Upsilon$ cross section, we calculate the effective cross section, \sigeff. 
The main sources of systematic uncertainty in the \sigeff~measurement are the estimates of the trigger efficiency and combinatorial background.
Based on Eq.~(\ref{eq:s_eff}) and upon the assumption~\cite{new_ref_dp} that $J/\psi+\Upsilon$ production has a negligible SP contribution, we obtain
\begin{eqnarray}
 \sigma_{\rm eff} 
= 2.2\pm 0.7 \mbox{\thinspace(stat)} \pm 0.9 \mbox{\thinspace(syst)} ~\mbox{mb}.
 \label{eq:eff_sigma_0}
\end{eqnarray}
The measured \sigeff~agrees with the result reported by the AFS Collaboration 
in the 4-jet final state~\cite{afs2} ($\approx 5$ mb) and D0 in the double $J/\psi$ final state~\cite{D0JJ} [$4.8 \pm 0.5 \mbox{\thinspace(stat)} \pm 2.5 \mbox{\thinspace(syst)} ~\mbox{mb}$].
However, it is lower than the CDF results
 in the 4-jet final state~\cite{cdf1} [$12.1^{+10.7}_{-5.4}$ mb] and $\gamma/\pi^0+3$-jet final state~\cite{cdf2} [$14.5 \pm 1.7\thinspace({\rm stat}) ^{+1.7} _{-2.3}\thinspace({\rm syst})$ mb];
the D0~\cite{D0_2} result in $\gamma + 3$-jet
events~\cite{D0_2} [$12.7 \pm 0.2\thinspace({\rm stat}) \pm 1.3\thinspace({\rm syst})$ mb];
 both ATLAS~\cite{atlas} [$15 \pm 3\thinspace({\rm stat}) ^{+5} _{-3}\thinspace({\rm syst})$ mb] and CMS~\cite{cms_w2j} [$20.7 \pm 0.8\thinspace({\rm stat}) \pm 6.6\thinspace({\rm syst})$ mb] results in the $W$+2-jet final state;
and the LHCb~\cite{lhcb_bbcc} [$18.0 \pm 1.3\thinspace({\rm stat}) \pm 1.2\thinspace({\rm syst})$ mb] result in $\Upsilon+$ open charm events.
The DP $J/\psi$+$\Upsilon$, double $J/\psi$, and 4-jet production are dominated by $gg$ initial states, whereas the
 $\gamma(W)$+jets events are produced predominantly by $q \bar q'$, and $qg$ processes.  The values of \sigeff~measured in different final
 state channels indicate that gluons occupy a smaller region of space within the proton than quarks. 
The pion cloud model~\cite{cloud_mpd} predicts a smaller average transverse size of the gluon distribution in a nucleon than that for quarks.

In conclusion, we have presented the first evidence of simultaneous production of prompt  $J/\psi$ and $\Upsilon$ ($1S,2S,3S$) mesons with a significance of 3.2 standard deviations.
The process is expected to be dominated by double parton scattering. The distribution of the azimuthal angle between the $J/\psi$ and $\Upsilon$ candidates is 
consistent with the double parton scattering predictions. Under the assumption of it being entirely composed of double parton scattering,
in the fiducial region of $p^\mu_T>2$ GeV and $|\eta^\mu|<2$ we measure the cross section $\sigma_{\rm{DP}}(J/\psi+\Upsilon)=27 \pm 9 \mbox{\thinspace(stat)} \pm 7 \mbox{\thinspace(syst)} ~\mbox{fb}$.
We also measure the single $J/\psi$ and estimate the single $\Upsilon$ ($1S,2S,3S$) production cross sections in the same fiducial region as the $J/\psi+\Upsilon$ cross section 
and find the effective cross section for this $gg$ dominated process to be $\sigma_{\rm eff} = 2.2\pm0.7\mbox{\thinspace(stat)}\pm0.9\mbox{\thinspace(syst)}$ mb, lower than the values found in the $q \bar q$ and $qg$ dominated 
double parton processes. This suggests that the spatial region occupied by gluons within the proton is smaller than that occupied by quarks.

%

We thank the staffs at Fermilab and collaborating institutions,
and acknowledge support from the
Department of Energy and National Science Foundation (United States of America);
Alternative Energies and Atomic Energy Commission and
National Center for Scientific Research/National Institute of Nuclear and Particle Physics  (France);
Ministry of Education and Science of the Russian Federation, 
National Research Center ``Kurchatov Institute" of the Russian Federation, and 
Russian Foundation for Basic Research  (Russia);
National Council for the Development of Science and Technology and
Carlos Chagas Filho Foundation for the Support of Research in the State of Rio de Janeiro (Brazil);
Department of Atomic Energy and Department of Science and Technology (India);
Administrative Department of Science, Technology and Innovation (Colombia);
National Council of Science and Technology (Mexico);
National Research Foundation of Korea (Korea);
Foundation for Fundamental Research on Matter (The Netherlands);
Science and Technology Facilities Council and The Royal Society (United Kingdom);
Ministry of Education, Youth and Sports (Czech Republic);
Bundesministerium f\"{u}r Bildung und Forschung (Federal Ministry of Education and Research) and 
Deutsche Forschungsgemeinschaft (German Research Foundation) (Germany);
Science Foundation Ireland (Ireland);
Swedish Research Council (Sweden);
China Academy of Sciences and National Natural Science Foundation of China (China);
and
Ministry of Education and Science of Ukraine (Ukraine).
%



\begin{thebibliography}{99}
\bibitem{coord}
Pseudorapidity is defined as $\eta = -\ln[\tan(\theta/2)]$, where $\theta$ is the polar angle with
respect to the positive $z$ axis along the proton beam direction, while the azimuthal angle $\phi$ is defined
with respect to the $x$ axis pointing away from the center of the Tevatron ring. The $y$ axis points upwards.
\end{thebibliography}
\end{document}